\documentstyle{article}
\input epsf

\begin{document} 

\begin{flushright}
SUSSEX-TH-98-021\\
PU-RCG-98/14\\
{\sf hep-th/9809105}\\
{\em to appear in Physics Letters B}
\end{flushright}

\begin{center} 
\Large
{\bf Axion perturbation spectra in string cosmologies}\\ 
\bigskip
\large 
E.~J.~Copeland$^{1}$, James E.~Lidsey$^{1,2}$ and  David Wands$^{3}$\\
\bigskip
\normalsize
$^{1}$Centre for Theoretical Physics, University of Sussex, Brighton,
BN1 9QJ, U.K.\\ 
\vspace{.2cm} 
$^{2}$Astronomy Centre, University of Sussex, Brighton, BN1 9QJ, U.K.\\ 
\vspace{.2cm} 
$^{3}$School of Computer Science and Mathematics, 
University of Portsmouth, Portsmouth PO1 2EG, U.K.
\end{center} 
\begin{abstract} 
We discuss the semi-classical perturbation spectra produced in the
massless fields of the low energy string action in a pre big bang type
scenario. Axion fields may possess an almost scale-invariant ($\Delta
n\approx0$) spectrum on large scales dependent upon the evolution of
the dilaton and moduli fields to which they are coupled. As an example
we calculate the spectra for three axion fields present in a truncated
type IIB model and show that they are related with at least one of the
fields having a scale-invariant or red ($\Delta n<0$) perturbation
spectrum. 
% XXXX
In the simplest pre big bang scenario this may be inconsistent with
the observed isotropy of the microwave background. 
More generally the relations between the perturbation
spectra in low energy string cosmologies should reflect the symmetries
of the theory.

\end{abstract}

Superstring theory is at present the best candidate for a theory
uniting gravity with the other fundamental forces. If it is the correct
description of our physical world, it must have important consequences
for models of the very early universe. Much
theoretical work is currently devoted to building models of
cosmological inflation driven by slow-rolling, self-interacting
scalar fields in the context of supergravity models. However, there are
problems inherent in such an approach due to the large masses for the
scalar fields that are generally introduced by supergravity
terms~\cite{Lyth98}.
A radically different cosmological scenario has been proposed by
Gasperini and Veneziano based on the low-energy vacuum solutions
derived from the generic superstring effective action
\cite{pBB}. Inflation in this pre big bang model is driven by the
kinetic energy of the fast-rolling dilaton field rather than any
interaction potential. A number of new problems appear in such a
scenario, most notably the graceful exit
problem~\cite{graceful}. There are also concerns
about the specific initial conditions required~\cite{tuning}.

The key test of all these models of the very early universe is the
spectrum of perturbations that they predict. In conventional slow-roll
inflation the only perturbation spectra usually generated are the
gravitational wave background and perturbations in a single
quasi-massless inflaton field. This naturally produces a nearly
scale-invariant spectrum of adiabatic density perturbations. By
contrast there are potentially many massless scalar fields in a pre
big bang string cosmology which each produce their own spectrum of
perturbations. The fast-rolling dilaton and moduli fields can only
yield a steep blue spectrum \cite{BrusteinET1995}. However, it was
recently realised that axion fields, which are always present in the
low-energy effective action \cite{effaction}, may have significantly
different spectral slopes due to their explicit coupling to the
dilaton and moduli fields \cite{CEW97,CLW97}.

For instance, the pseudo-scalar axion field, $\sigma_1$, whose
gradient is dual to the Neveu/Schwarz-Neveu/Schwarz (NS-NS) three-form
field strength in four-dimensions \cite{Sdual}, can have a spectral
tilt in the range from $3-2\sqrt{3}\simeq-0.46 \le \Delta n_1 \le 3$,
depending upon the evolution of the dilaton and moduli
fields~\cite{CEW97,CLW97}.  In highly symmetrical cases the spectrum
becomes scale-invariant, $\Delta n_1=0$ \cite{b}. Durrer {\em et al.}
\cite{Durrer} have noted that such a spectrum may provide a novel
scenario for structure formation induced by seed perturbations. The
massless axion field can yield 
% XXXX
an almost scale-invariant spectrum of density
perturbations at horizon crossing whose amplitude is fixed by the
string coupling constant at the end of the pre big bang era,
% XXXX
$\delta\rho/\rho\approx e^\varphi \approx 10^{-2}$.
% XXXX
Hence a slightly ``blue'' spectrum, with $\Delta n_1>0$, may be
consistent with observations of the microwave
background anisotropies.  
More detailed analyses are required to examine whether this model of
structure formation can compete with the successful model based around
conventional inflation \cite{LiddleLyth}, but this example emphasises
the potential challenge to the standard picture raised by the pre big
bang scenario as well as the importance of perturbation spectra in
testing these ideas.

Thus far, calculations of axion perturbations have been restricted to
simple axion-dilaton systems \cite{CEW97,CLW97,b,BH98,BGV98}.  In this
paper we investigate another important feature of the pre big bang
model which has received little attention to date. The various
(pseudo--) scalar axion fields present in low energy effective actions
have different perturbation spectra due to their different couplings
to the dilaton and moduli fields. In general, however, these numerous
fields are coupled to the {\em same} dilaton and moduli which will
lead to distinctive relations between the corresponding perturbation
spectra.  As an example, we explore such a relation within the context
of a triple axion system derived from the type IIB superstring reduced
to four-dimensions~\cite{CLWiib}.  This model has recently been
studied in the context of homogeneous cosmological
backgrounds~\cite{CLWiibcosmo}.

The type IIB superstring contains a dilaton, a graviton and a 
two--form potential in the NS--NS sector of the theory, 
together with a second two--form potential and an axion field, 
$\sigma_3$,  in the Ramond--Ramond (RR) sector. (There is also 
a four--form in the RR sector, but this can be consistently set 
to zero). A 4-D action may be derived by compactifying the 
10-D spacetime on a 6-D Ricci-flat internal space so that 
\begin{equation}
\label{ansatz}
ds_{10}^2 = e^{\varphi(x)} \tilde{g}_{\mu\nu} (x) dx^{\mu} dx^{\nu} 
+e^{y(x)/\sqrt{3}} g_{ab} dX^a dX^b  ,
\end{equation}
where $\varphi$ is the 4-D dilaton and $y$
describes the volume of the internal space and is the only modulus
field considered. We have included the conformal factor $e^\varphi$ in
our definition of the 4-D external metric $\tilde{g}_{\mu\nu}$ in
order to work in the 4-D Einstein frame, where 
the dilaton field is minimally coupled to gravity.
In four dimensions the three-form field strengths from the NS-NS and
RR sectors are dual to the gradients of two
pseudo-scalar axion fields $\sigma_1$ and $\sigma_2$. The third axion
field is the RR axion already present in the 10-D theory.
In this dual formulation it can be shown that 
the equations of motion for the fields 
follow from an effective action \cite{CLWiib}:
\begin{eqnarray}
\label{solitonicaction}
S&=&{1\over2\kappa^2} 
\int d^4 x \sqrt{-\tilde{g}} \left[ \tilde{R} -\frac{1}{2} 
\left( \tilde{\nabla} \varphi \right)^2 - {1\over2} \left( 
\tilde{\nabla} y \right)^2  
 \right. \nonumber \\
&& \left. 
-\frac{1}{2} e^{\sqrt{3}y +\varphi} \left( \tilde{\nabla} \sigma_3 \right)^2 
-\frac{1}{2} e^{-\sqrt{3}y +\varphi} \left( \tilde{\nabla} \sigma_2 \right)^2
% \right. \nonumber \\
%&& \left. 
-\frac{1}{2} 
e^{2\varphi} \left( \tilde{\nabla} \sigma_1 -\sigma_3 
\tilde{\nabla} \sigma_2 \right)^2 
\right]  \ .
\end{eqnarray}
where $\kappa^2=8\pi/m_{\rm Pl}^2$.
This describes a non-linear sigma model in Einstein gravity where 
the scalar fields parametrise an SL(3,R)/SO(3) 
coset~\cite{CLWiib}. The global symmetries 
of the action include the SL(2,Z) `S-duality' of
the original 10-D action \cite{s} and a Z$_2$ `T-duality' corresponding to
invariance under $y\to-y$ \cite{meissner}.

We assume that the external four dimensional spacetime is described by a
flat Friedmann-Robertson-Walker (FRW) metric with the line element 
\begin{equation}
\label{FRW}
d\tilde{s}^2 = \tilde{a}^2(\eta) \left\{ -d\eta^2 +
\delta_{ij}dx^idx^j \right\}  \, ,
\end{equation}
where $\eta$ is the conformally invariant time coordinate and
$\tilde{a} (\eta)$ is the scale factor. FRW solutions with non-zero spatial
curvature can also be found~\cite{CLW94,CLWiibcosmo}.
The familiar field equations of general relativity apply in the
4-D Einstein frame 
and the combined stress-energy tensor for homogeneous massless fields reduces
to that for a perfect fluid with a maximally stiff equation of
state~\cite{Tab+Taub,MW95}, i.e., with pressure equal to energy density.
This leads to the simple solution for the scale factor
\begin{equation}
\label{Esf}
\tilde{a} = \tilde{a}_* |\eta|^{1/2} \, .
\end{equation}

The dilaton-moduli-vacuum solutions are monotonic power-law solutions
\begin{eqnarray}
\label{dilphi}
e^\varphi & = & e^{\varphi_*} |\eta|^{\sqrt{3}\cos\xi}
 \, ,\\
\label{dilbeta}
e^y & = & e^{y_*} |\eta|^{\sqrt{3}\sin\xi} \, ,
\end{eqnarray}
where the integration constant $\xi$ determines the relative rate of
change of the effective dilaton and internal volume respectively.
If stable compactification has occurred and the volume of the
internal space is fixed,  we have $\sin\xi=0$.

In order to understand the perturbation spectra produced in different
fields it is revealing to look at conformally related metrics,
$g_{\mu\nu}\to\Omega^2g_{\mu\nu}$. If the conformal factor $\Omega^2$ is
itself homogeneous, the transformed metric remains a FRW metric
but with scale factor $a\to\Omega a$ and proper time 
$t\to\int\Omega dt$. A finite proper
time interval in one frame does not necessarily coincide with a finite
proper time in another frame and, in particular, we shall see that what 
seems to be a 
singular evolution in one frame may appear to be non-singular in
another frame.

In the original string frame the scale factor evolves as
\begin{equation}
\label{dila}
a  \equiv   e^{\varphi/2} \tilde{a} = a_* |\eta|^{(1+\sqrt{3}\cos\xi)/2}
 \, ,
\end{equation}
and there is an accelerated expansion in this
frame for $\cos\xi<-1/\sqrt{3}$ if $\eta<0$.
In the Einstein frame we see that $\eta\to0^-$ always
corresponds to a collapsing universe with $\tilde{a}\to0$. However,
for any value of $\xi$ this fulfils one definition of inflation,
namely, that the comoving Hubble length
($|d\tilde{a}/d\tilde{t}|^{-1} =
|\tilde{a}/\tilde{a}'|=2|\eta|$) decreases with
time~\cite{Deflation}. A given comoving scale that starts
arbitrarily far within the Hubble scale in either conformal frame at
$\eta\to-\infty$ inevitably becomes larger than the Hubble scale in
that frame as $\eta\to0^-$. 
This allows one to produce perturbations in
the fields on scales much larger than the
present Hubble scale from quantum fluctuations in flat-spacetime at
earlier times. 

Because the dilaton and moduli are both minimally coupled to the
Einstein metric,  
the field equations for their linearised scalar perturbations are given by 
\cite{BrusteinET1995,Hwang,Giovannini}
\begin{eqnarray}
\label{dphieom}
\delta\varphi'' + 2\tilde{h}\delta\varphi' + k^2\delta\varphi
 & = & 0
\\
\label{dbetaeom}
\delta y'' +2\tilde{h}\delta y' +k^2\delta y
 & = & 0
\end{eqnarray}
where a prime denotes differentiation with respect to conformal time,
the comoving Hubble rate in the Einstein frame is given by
$\tilde{h}\equiv\tilde{a}'/\tilde{a}$ and $k$ is the comoving
wavenumber. (Perturbations in the gravitational field obey a similar
field equation \cite{GW}).  The singular evolution of the metric as
$\eta \to 0^-$ implies that their perturbation spectra grow
dramatically on shorter wavelengths that leave the Hubble radius close
to the singularity.  This leads to steep blue spectra with spectral
tilt $\Delta n=3$~\cite{BrusteinET1995} which leaves effectively no
perturbations in these fields on large (astronomically observable)
scales in our present universe.

On the other hand, axion fields' kinetic terms retain a non-minimal
coupling to the dilaton or moduli fields in the Einstein frame.
This non-minimal coupling can be eliminated by a
conformal transformation to an alternative conformally related
metric, which we will refer to as the corresponding axion
frame. For the NS-NS axion this is given by~\cite{CEW97}
$\bar{g}_{(1) \mu\nu}=e^{2\varphi}\tilde{g}_{\mu\nu}$ 
and hence
\begin{equation}
\label{bara}
\bar{a}_1 \equiv e^\varphi\tilde{a} \, .
\end{equation}
The NS-NS axion field is a minimally coupled massless scalar field in
this frame and thus axionic particles follow null geodesics with
respect to this metric.

More generally, for the three axion fields in the truncated type IIB 
action given in Eq.~(\ref{solitonicaction}) we can define
\begin{equation}
\bar{a}_i^2 = \Omega_i^2 \tilde{a}^2 \,,
\end{equation}
where the conformal factor
\begin{equation}
\label{Omega}
\Omega_i^2 = \left\{
\begin{array}{ll}
e^{2\varphi} & {\rm for}\ \sigma_1 \\
e^{\varphi-\sqrt{3}y} & {\rm for}\ \sigma_2 \\
e^{\varphi+\sqrt{3}y} & {\rm for}\ \sigma_3
\end{array}
\right. \,,
\end{equation}
reflects the differing couplings of the axion fields to the dilaton
and moduli in the Lagrangian.

Although conformally related to the string and Einstein frames, the
metric ``seen'' by the axions may behave very differently. 
In terms of conformal time, the axionic scale factors for the
dilaton-moduli-vacuum solutions given by Eqs.~(\ref{dilphi}) and~(\ref{dila})
evolve as
\begin{equation} 
\label{axionsf}
\bar{a}_i = \bar{a}_{*i} |\eta|^{r_i+(1/2)} \,  
\end{equation} 
where
\begin{equation}
r_i = \left\{
\begin{array}{ll}
\sqrt{3} \cos\xi & {\rm for}\ \sigma_1 \\
\sqrt{3} \cos(\xi+\pi/3) & {\rm for}\ \sigma_2 \\
\sqrt{3} \cos(\xi-\pi/3) & {\rm for}\ \sigma_3
\end{array}
\right. \,.
\end{equation}
The proper time in the axion frame is given by 
\begin{equation}
\bar{t}_i\equiv\int\bar{a}_i\,d\eta \sim |\eta|^{r_i+(3/2)} \, , 
\end{equation} 
so it takes an infinite proper time to reach $\eta=0$ for $r_i\leq-3/2$
and the scalar curvature for the axion metric,
$\bar{R}_i\sim\bar{t}_i^{-2}$, vanishes as $\eta\to0$. However, these
same dilaton-moduli-vacuum solutions then reach $\eta\to\pm\infty$ in a finite
proper time where $\bar{R}_i$ diverges. Because the conformal factor
diverges as $|\eta|\to0$ it stretches out the curvature singularity in the
string metric into a non-singular evolution in the axion frame. But since 
$\Omega^2=e^\varphi\to0$ as $\eta\to\pm\infty$ the non-singular 
asymptotic behaviour in this limit in the Einstein or string frames
gets compressed into a curvature singularity in the
axion frame.

In terms of the proper time in the axion frame we have
\begin{equation}
\bar{a}_i = \bar{a}_{*i} \left( {\bar{t}_i\over\bar{t}_{*i}}
 \right)^{(1+2r_i)/(3+2r_i)}
\end{equation}
For $r_i<-3/2$ we have conventional power-law inflation (not
pole-inflation) with $\ln \bar{a}_i \sim \bar{p}_i  \ln
\bar{t}_i$, where
$\bar{p}_i=1+[2/(-2r_i-3)]>1$. This has important
consequences for the tilt of the power spectrum of semi-classical
perturbations in the axion field produced on large scales.

The field equations for the linearised scalar perturbations 
in the axion fields 
are~\cite{CEW97} 
\begin{equation}
\label{dsigmaeom}
\delta\sigma_i''+2\bar{h}_i\delta\sigma_i' + k^2\delta\sigma_i
=  0 
\end{equation}
where the comoving Hubble rate in the axion frame for 
each field is given by
\begin{equation}
\bar{h}_i \equiv {\bar{a}_i'\over\bar{a}_i}
 = {\tilde{a}'\over\tilde{a}} + {\Omega_i' \over \Omega_i} \ .
\end{equation}
The canonically normalised axion field perturbations are given 
by~\cite{Mukhanov88,CEW97} 
\begin{equation}
\label{defv}
v_i \equiv {1\over\sqrt{2}\kappa} \bar{a}_i\delta\sigma_i \, 
\end{equation}
and the equation of motion given in Eq.~(\ref{dsigmaeom}) can be
re-written in terms of $v_i$ as
\begin{equation}
\label{vpp}
v_i'' + \left( k^2 - {\bar{a}_i''\over\bar{a}_i} \right) v_i = 0 \, .
\end{equation}
In the terminology of Ref.~\cite{BGV98}, the pump field $S$ for the
perturbations in each axion field is given by the square of the scale
factor in the corresponding conformal frame, $S_i=\bar{a}_i^2$.
After inserting the power-law solution for the axion frame scale 
factor given in Eq. (\ref{axionsf}), we find that these equations give 
the general solutions 
\begin{equation}
\label{vsol}
v_i  = 
 |k\eta|^{1/2} \left[ 
  v_+ H_{\mu_i}^{(1)}(|k\eta|) + v_- H_{\mu_i}^{(2)}(|k\eta|) 
 \right] \, ,
\end{equation}
where $H^{(j)}_{\mu_i}$ are Hankel functions of order $\mu_i=|r_i|$. 

For pre big bang solutions, i.e., $\eta<0$, we can normalise modes 
on small scales at early times when all the modes are far inside the
Hubble scale, $k\gg|\eta|^{-1}$. They can be assumed to be in
the flat-spacetime vacuum\footnote{It is interesting to note that in
conventional inflation we have to assume that this result for a
quantum field in a classical background holds at the Planck
scale. Here, however, the normalisation is done in the zero-curvature
limit in the infinite past.}. 
Allowing only positive frequency modes in the flat-spacetime vacuum
state at early times requires that
\begin{equation}
\label{shortwave}
v_i \to {e^{-ik\eta} \over \sqrt{2k}} 
\end{equation}
as $k\eta\to-\infty$ \cite{BirrellDavies}. This gives
\begin{equation}
\label{vplus}
v_+= e^{i(2\mu_i+1)\pi/4} {\sqrt{\pi} \over 2\sqrt{k}} \, , \qquad v_-=0 \, 
\end{equation}
and hence we have
\begin{equation}
\label{pBBdsigma}
\delta\sigma_i = \kappa \sqrt{{\pi \over 2k}} e^{i(2\mu_i+1)\pi/4}
{\sqrt{-k\eta} \over \bar{a}} H_{\mu_i}^{(1)}(-k\eta) \, .
\end{equation}

Just as in conventional inflation, this
produces perturbations on scales far outside the horizon,
$k\ll|\eta|^{-1}$, at late times, $\eta\to0^-$.
The power spectrum for perturbations is conventionally denoted by
\begin{equation}
{\cal P}_{\delta x} \equiv {k^3\over2\pi^2} |\delta x|^2 \, ,
\end{equation}
and thus for modes far outside the horizon ($-k\eta\to0$) 
we have\footnote
{When $r=0$ the dilaton remains constant and the axion
frame and Einstein frame coincide, up to a constant factor. Thus, the
axion spectra behave in the same way 
as those of the dilaton and moduli fields. 
The late time evolution in this case is logarithmic with respect
to $-k\eta$ \cite{BrusteinET1995}.}
\begin{equation}
\label{pBBsigma}
{\cal P}_{\delta\sigma_i} = 2\kappa^2 \left( {C(\mu_i) \over 2\pi} \right)^2
 {k^2\over\bar{a}^2} (-k\eta)^{1-2\mu_i} \, ,
\end{equation}
where the numerical coefficient
\begin{equation}
C(\mu_i) \equiv {2^{\mu_i}\Gamma(\mu_i) \over 2^{3/2}\Gamma(3/2)} 
\end{equation}
approaches unity for $\mu_i=3/2$.

The expression for the axion power spectrum can be written in terms of
the field perturbation when each mode crosses outside the horizon 
($k\eta_c=-1$):
\begin{equation}
\left. {\cal P}_{\delta\sigma_i} \right|_c
 = 2\kappa^2 \left[{C(\mu_i)\over r_i+(1/2)}\right]^2
 \left( {\bar{H}_i \over 2\pi} \right)_c^2 \, ,
\end{equation}
where $\bar{H}_i|_c$ is the Hubble rate in the axion frame\footnote
{The Hubble rate in the axion frame can be written as 
$\bar{H}_i =  (1+2r_i) \Omega^{-1}_i \tilde{H}$, 
where the conformal factor $\Omega^2_i$ is given in Eq.~(\ref{Omega}) and 
$\tilde{H}$ is the expansion parameter in the Einstein frame.}
when $k\eta_c=-1$.
This is the power spectrum for a massless scalar field during
power-law inflation which approaches the famous result\footnote
{The factor $2\kappa^2$ arises due to our dimensionless definition of
$\sigma_i$.}
${\cal P}_{\delta\sigma_i}|_c=2\kappa^2(\bar{H}_i/2\pi)_c^2$
 as $r_i\to-3/2$, this critical case arising when 
the expansion in the axion frame becomes exponential. 

The amplitude of the power spectra at the end of the pre big bang
phase can be written as
\begin{equation}
\label{Pdsend}
\left. {\cal P}_{\delta\sigma_i} \right|_s
 = 2\kappa^2 \left[{C(\mu_i)\over r_i+(1/2)}\right]^2
 \left( {\bar{H}_i \over 2\pi} \right)_s^2
 \left( {k\over k_s} \right)^{3-2\mu_i}\, ,
\end{equation}
where $k_s$ is the comoving wavenumber of the scale just leaving the
Hubble radius at the end of the pre big bang phase, $k_s\eta_s=-1$. 
The subsequent evolution of these perturbations may depend upon the
nature of the exit from the pre big bang ($\eta<0$) to the post big
bang ($\eta>0$). The simplest assumption\footnote
{This simple assumption was verified in the specific scenarios
investigated in Ref.~\cite{b}.}
is that the modes remain
frozen-in on large scales ($|k\eta|\ll1$) in which case the massless
axion perturbations contribute an energy density
\begin{equation}
\tilde\rho_i \sim
 \Omega_i^2 {k^2\over\tilde{a}^2} {\left. {\cal P}_{\delta\sigma_i}
 \right|_s \over 2\kappa^2}  
= C^2(\mu_i) {k^2\over\tilde{a}^2} \left( {\tilde{H} \over 2\pi} \right)_s^2 
 \left( {k\over k_s} \right)^{3-2\mu_i}\, ,
\end{equation}
in the Einstein frame. 
We note that although the amplitude of the perturbations in each axion
field depends upon the conformal factor $\Omega_i^2$, the effective
energy density in the Einstein frame for perturbations with $k\sim
k_s$ of a massless axion field are independent of $\Omega_i^2$. Thus, 
the amplitude of density perturbations on larger scales depends only
upon the spectral tilt.
% XXXX
Durrer {\em et al.} have noted that a spectrum for a massless axion, 
slightly tilted towards smaller scales,
may be consistent with the
observed amplitude of anisotropies in the cosmic microwave background
with $\Delta T/T\sim (\tilde\rho_i)_{k=aH}/\rho_{\rm crit}
\sim\kappa^2 \tilde{H}_s^2(k/k_s)^{3-2\mu_i}$ for 
$\kappa^2\tilde{H}_s^2\sim e^\varphi\sim 10^{-2}$ \cite{Durrer}.  

The spectral tilt of the perturbation spectra is given by
\begin{equation}
\label{specindex}
\Delta n_i \equiv {d\ln{\cal P}_{\delta\sigma_i} \over d\ln k}
\end{equation}
The spectral tilt for each 
of the fields follows from Eq.~(\ref{pBBsigma}), and are shown in Figure~1. 
They take the values 
\begin{equation}
\Delta n_i = 3 - 2\mu_i = 3 - 2\sqrt{3}|\cos(\xi-\xi_i)|
\end{equation}
where 
\begin{equation}
\xi_i = \left\{
\begin{array}{ll}
0 & {\rm for}\ \sigma_1 \\
 -\pi/3 & {\rm for}\ \sigma_2\\
\pi/3 & {\rm for}\ \sigma_3
\end{array}
\right. \,.
\end{equation}
The tilts depend crucially upon the value of $\mu_i$. 
% XXXX ``classic Harrison-Zel'dovich'' removed!
The spectrum becomes scale-invariant spectrum in the limit
$\mu_i\to3/2$. The lowest possible value of the spectral index for any
of the axion fields is $3-2\sqrt{3}\simeq-0.46$.
Requiring conventional power-law inflation, rather than pole
inflation, in the axion frame, guarantees a negatively tilted spectrum
($\Delta n_i<0$)\footnote
% XXXX ``for modes outside the horizon'' added to clarify point raised
% by Veneziano
{Note that although the power spectrum for axion perturbations
diverges on large scales for $\Delta n_i<0$, the energy density for
modes outsde the horizon is proportional to $k^2{\cal
P}_{\delta\sigma_i}$ and this remains finite.}. 

\begin{figure}[t]
\centering 
\leavevmode\epsfysize=5cm \epsfbox{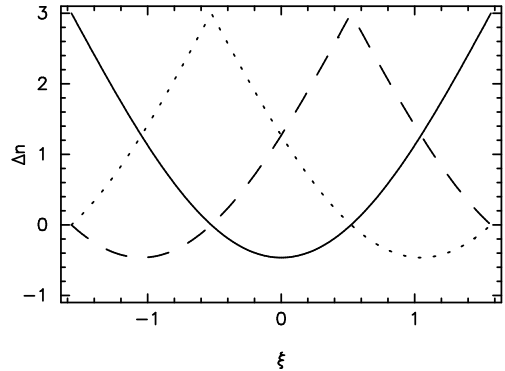}\\ 
\caption[Spectral indices]{\label{ni} Spectral tilts $\Delta n_i$ for three
axion fields' perturbation spectra in truncated type IIB action as a
function of integration constant $\xi$ in pre big bang solutions. 
The solid line corresponds to $\Delta n_1$, the dotted line to $\Delta
n_2$ and the dashed line to $\Delta n_3$.}
\end{figure}

All three spectral indices for the axion fields in the truncated
type IIB model which we have considered are determined by the single
integration constant $\xi$.  In particular, we find that one of the
axion fields {\em always} has a red spectrum ($\Delta n_i<0$) while
the other two spectra are blue ($\Delta n_i>0$), except in the
critical case $|\cos\xi|=\sqrt{3}/2$, where two of the spectra are
scale-invariant and only one is blue.  This provides an example of the
important phenomenological role that the RR sector of string theory
can play in cosmological solutions~\cite{RRobs}.

More generally, the axion perturbation spectra can have different
spectral indices, but in a given string model there is a {\em
specific} relationship between them. This follows as a direct
consequence of the symmetries of the effective action. These
symmetries relate the coupling parameters between the various fields
and are manifested in the spectra.  Such perturbation spectra could
provide distinctive signatures of the early evolution of our universe.
The analysis presented above should be applicable to a wide class of
non-linear sigma models coupled to gravity.  In such models, the
couplings between the massless scalar fields are specified by the
functional form of the target space metric. These couplings determine
the appropriate conformal factors analogous to those in
Eq.~(\ref{Omega}) that leave the fields minimally coupled and it is
the evolution of these couplings that directly determine the scale
dependence of the perturbation spectra.

% XXXX
Because large symmetry groups which include SL(3,R) are ubiquitous in
supergravity theories obtained from compactification of higher
dimensional theories \cite{cj}, our result raises a serious challenge
for the pre big bang scenario. We have shown that at least one
massless axion field in an SL(3,R) non-linear sigma model will have a
negatively tilted spectrum. In the scenario considered by Durrer et
al.~\cite{Durrer} the amplitude of density perturbations at horizon
crossing is determined by the string scale at the end of the pre big
bang era, $\delta\rho/\rho\sim e^\varphi$. Only a positively tilted
spectrum can be consistent with the usual string scale,
$e^\varphi\sim10^{-2}$ if density perturbations on larger scales are
to remain compatible with the isotropy of the microwave background
sky. This assumes that the axion remains massless. One would na\"\i
vely expect that the introduction of a mass for the axion field would
only make matters worse. One possible way out, would be for the axion
to develop a periodic potential in which case the axion might
contribute a large fraction of the dark matter in our universe, but
the large field fluctuations might lead to only small density
fluctuations~\cite{BH98b}.

\vspace{.1in}

{\bf Acknowledgements}

\vspace{.1in}

We are grateful to Andrew Liddle for useful discussions, 
% XXXX
and to Gabriele Veneziano for drawing our attention to
the problems posed by the negatively tilted axion spectra. 
EJC and JEL are supported by the Particle Physics and Astronomy 
Research Council (PPARC), UK.

\end{document}